# Deep learning to estimate the physical proportion of infected region of lung for COVID-19 pneumonia with CT image set

Wei Wu[1], Yu Shi[1], Xukun Li[2], Yukun Zhou[2], Peng Du[2], Shuangzhi Lv[3], Tingbo Liang[4], Jifang Sheng[1]

1. State Key Laboratory for Diagnosis and Treatment of Infectious Diseases, National Clinical Research Center for Infectious Diseases, Collaborative Innovation Center for Diagnosis and Treatment of Infectious Diseases, the First Affiliated Hospital, School of Medicine, Zhejiang University, Hangzhou 310003, China.
2. Artificial Intelligence Lab, Hangzhou AiSmartVision Co., Ltd., Hangzhou, Zhejiang 310012, People's Republic of China
3. Department of Radiology The First Affiliated Hospital, School of Medicine, Zhejiang University, Hangzhou, Zhejiang 310003, People's Republic of China
4. Zhejiang Provincial Key Laboratory of Pancreatic Disease, Department of Hepatobiliary and Pancreatic Surgery, Innovation Center for the Study of Pancreatic Diseases, the First Affiliated Hospital, School of Medicine, Zhejiang University, Hangzhou 310003, China.

Correspondence to Wei Wu, PhD, MD
E-mail: 1198042@zju.edu.cn
ORCID: 0000-0003-4657-4088

**Abstract**

Utilizing computed tomography (CT) images to quickly estimate the severity of cases with COVID-19 is one of the most straightforward and efficacious methods. However, due to the three-dimensional structure and ambiguous edge of infected regions, it may be difficult to diagnose quantitatively for clinical physicians and radiologists.

Two tasks were studied in this present paper. One was to segment the mask of intact lung in case of pneumonia. Another was to generate the masks of regions infected by COVID-19. The masks of these two parts of images then were converted to corresponding volumes to calculate the physical proportion of infected region of lung.

A total of 129 CT image set were herein collected and studied. The intrinsic Hounsfiled value of CT images was firstly utilized to generate the initial dirty version of labeled masks both for intact lung and infected regions. Then, the samples were carefully adjusted and improved by two professional radiologists to generate the final training set and test benchmark. Two deep learning models were evaluated: UNet and 2.5D UNet. For the segment of infected regions, a deep learning based classifier was followed to remove unrelated blur-edged regions that were wrongly segmented out such as air tube and blood vessel tissue etc.

For the segmented masks of intact lung and infected regions, the best method could achieve 0.972 and 0.757 measure in mean Dice similarity coefficient on our test benchmark. As the overall proportion of infected region of lung, the final result showed 0.961 (Pearson's correlation coefficient) and 11.7% (mean absolute percent error).

The instant proportion of infected regions of lung could be used as a visual evidence to assist clinical physician to determine the severity of the case. Furthermore, a quantified report of infected regions can help predict the prognosis for COVID-19 cases which were scanned periodically within the treatment cycle.

Keywords: COVID-19, deep learning, coronavirus pneumonia, CT image set

# Introduction

Coronavirus disease 2019 or COVID-19 had become a worldwide pandemic and caused great public health problems[1-3]. COVID-19 cases can be divided into light, moderate, severe, and extremely severe types from physicians' perspective. Patients of the latter two types exhibited to have a higher intensive care unit (ICU) rates as well as the death rates [4,5] compared with the other two types. It is therefore essential to identify severe and extremely severe patients as early as possible.

In the Diagnosis and Treatment Protocol for COVID-19 (version 7) [6] released by National Health Commission of China, the clinical characteristics of severe cases include: the decreased lymphocyte count, increased level of inflammatory factors, and rapid development of volume for infected regions on CT images. One patient should be classifiled and treated as severe case if the volume of infected region would be increased more than 50% within 48 hours. Therefore, continuously monitoring the volume of infected regions may provide valid evidence to predict the prognosis for COVID-19 patients.

However, with infected regions in the lung, the Hounsfiled Unit (HU) value of the lesion regions would be difficult to distinguish with healthy tissues. The infected regions may illustrate as a mist of blur-edged cloud or adhere together with normal tissues on CT images. It would be effort costly for a professional radiologist or physician to separate these lesion regions from healthy lung parenchyma. Furthermore, a set of CT images usually consisted of dozens or hundreds of lung images, which made it almost impossible to analyze the lesion regions quantitatively over the images manually. Therefore, it was urgent to find out an automatic method to estimate the proportion of infected region of lung for COVID-19 from chest CT scans.

To date, several researches had concentrated on the deep learning based models for diagnosing COVID-19. Some studies [7-10] demonstrated that COVID-19 can be distinguished from other types of pneumonia with good accuracy. Compared with the classifications models, the annotation of CT image samples is highly significant and much more time-consuming in the training of segmentation models for the intact lung as well as infected regions. Shan et al. [11] adopted the human-in-the-loop strategy to iteratively update the annotation of their training samples. Liu et al. [12] synthesized part of their training and test dataset with Generative Adversarial Network (GAN). They achieved 0.706 measured in mean Dice similarity coefficient (m-Dice) for the segmentation of infected regions and 0.961 (Pearson correlation coefficient) for the total percent volume of lung parenchyma that was affected by disease. Ma et al. [13] annotated 20 sets of COVID-19 CT images and utilized previous available lung dataset such as lung cancer to assist the segmentation. Yan et al. [14] also investigated the segmentation of infected regions due to COVID-19. They employed a team of six annotators with deep radiology background and proficient annotating skills to label the areas and boundaries of the intact lung and infection regions due to COVID-19. A feature variation block in the segmentation of infected regions was introduced, which could better differentiate the diseased area from the lung. Furthermore, they used the more effective progressive atrous spatial pyramid pooling in the feature extraction stage as well. The optimum m-Dice achieved in their studies for entire lung and infected regions were 0.987 and 0.726 respectively. These studies suffered from the tremendous effort to label the training samples as well as the relatively low accuracy measured in m-Dice.

In this study, we try to establish a fully automatic deep learning system to estimate the physical proportion of infected region of lung for COVID-19 pneumonia with CT image set. The main contribution of this paper can be summarized as:

First, the HU value of CT images was utilized as threshold to generate the initial "dirty" version of labeled masks for both intact lung and infected regions. These first round labeled samples significantly alleviated the labor expenses of annotation compared with start from scratch. Then these preliminary samples were revised and improved by two professional radiologists to generate the final training set and test benchmark.

Second, it was observed from our experiment that a certain number of blur-edged healthy structures, which had similar appearance as infected regions, were likely to be identified incorrectly. These kind of healthy tissues included air tube, blood vessel, and blur region of lung at the border etc. Therefore, a deep learning based classifier was employed to further clarify candidate regions that could effectively increase the accuracy of the final segmentation results. In addition, the proposed classifier was much easier to be trained compared with pixel-level segmentation models.

## Materials and methods

### *Study Dataset*

A total of 129 transverse–section CT samples were collected, including 105 from 105 patients (mean age 51 years; 58 [55.2%] male patients) with COVID-19 from the First Affiliated Hospital of Zhejiang University, from January 19 to March 31, 2020. Every COVID-19 patient was confirmed with reverse transcription polymerase chain reaction (RT–PCR) kit, and cases with no image manifestations on the chest CT images were excluded. There were 80 (62.0%) COVID-19 from light to moderate types, and the remaining 49 (38.0% ) cases from severe to extremely severe types respectively. All CT imaging was in the format of digital imaging and communications in medicine (DICOM) with 5mm thickness between slices.

The study was approved by the ethics committee of the First Affiliated Hospital, School of Medicine, Zhejiang University and all research was performed in accordance with relevant guidelines and regulations. All participants and/or their legal guardians signed the informed consent form prior to commencing the study.

A total of 108 CT samples (83.7%) were used for training and validation datasets and the remaining 21 CT sets (16.3%) were used as a test benchmark.

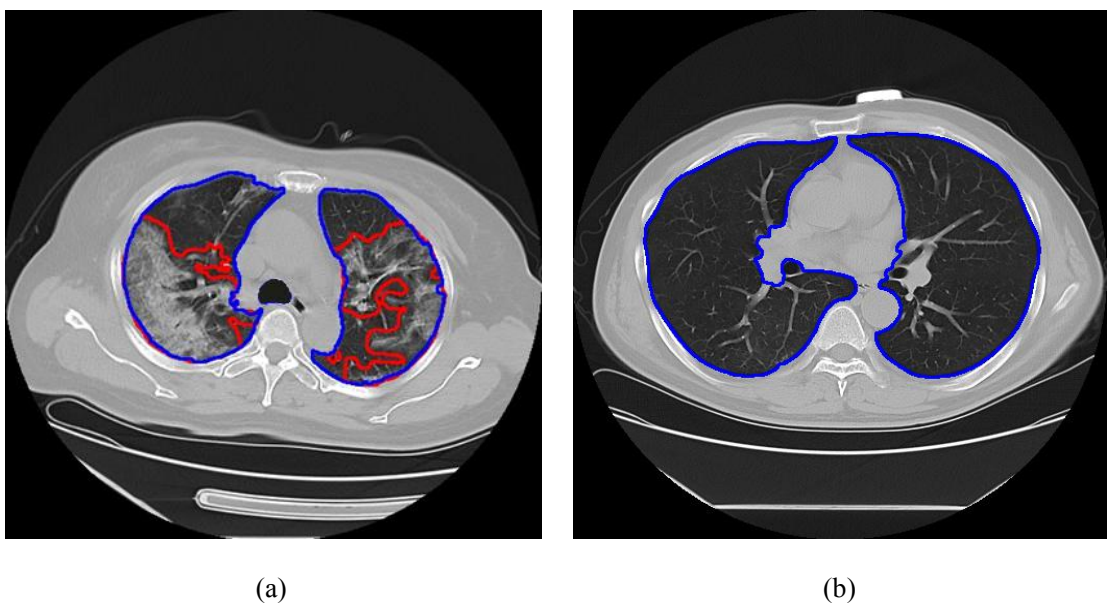

(a) (b)

**Figure 1**. Typical labeled CT images: (a) CT image with pneumonia; (b) CT image without pneumonia. The fields within the blue line denote the masks for the intact lung and those within the red line represent the masks for infected regions.

*Process*

Figure 2 showed the whole diagnostic process of COVID-19 report generation in this study. As the digital gray scale image had the pixel value ranging [0, 255], the raw data of CT were converted from HU to the aforementioned values interval accordingly. The HU data matrix was clipped within [–1200, 600] (any value beyond this was set to –1200 or 600 accordingly) and then linearly normalized to [0, 255] to fit into the digital image format for further processing. Next, the infected regions and the intact lung were segmented separately to achieve corresponding masks. For the differentiating of infected regions, a deep learning based classifier was utilized to remove unrelated masks that were wrongly distinguished. Finally, the volumes of two parts were calculated according to the masks achieved in above-mentioned steps, and achieve the proportion of infected regions in lung.

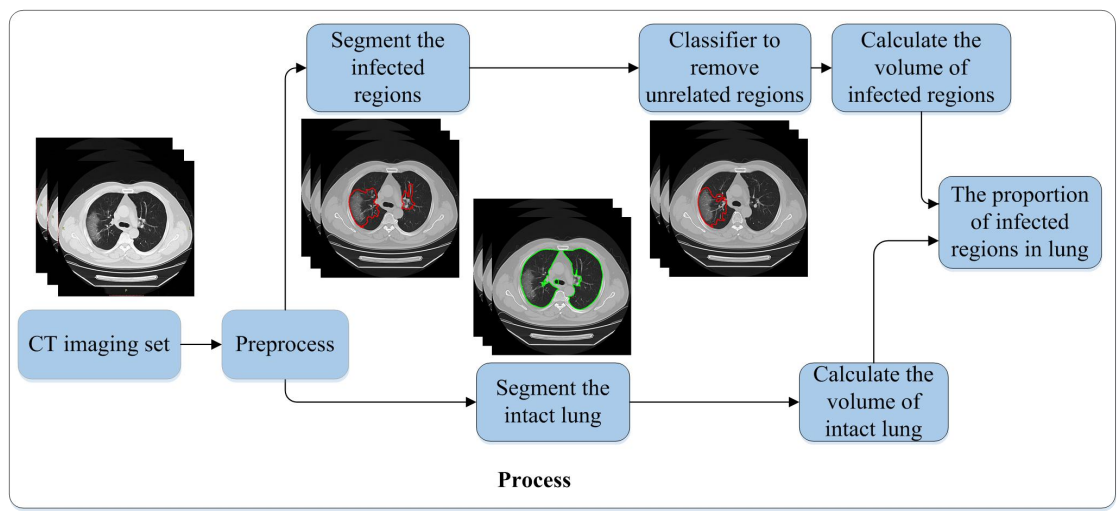

**Figure 2**. Study flow chart.

*Evaluation criteria*

The performance of the proposed method was evaluated using the Dice similarity coefficient (Dice), measuring the similarity between the ground truth and the prediction score maps. It is calculated as follows:

$$Dice(A, B) = \frac{2|A \cap B|}{|A| + |B|} \times 100\% \quad (1)$$

where A is the volume of the segmented lesion region and B denotes the ground truth. The mean Dice (m-Dice) of the whole test benchmark was used to evaluate the final outcomes. Two ground truth masks were used in this study: ground truth for intact lung and ground truth for infected regions. The proportion of infected regions of lung (PoIR) was given by:

$$PoIR = \frac{\text{volume of infected regions}}{\text{volume of intact lung}} \times 100\% \quad (2)$$

Pearson's correlation coefficient was used to evaluate the correlation of two variables:

$$r = \frac{N \sum_i x_i y_i - \sum_i x_i \sum_i y_i}{\sqrt{N \sum_i x_i^2 - (\sum_i x_i)^2} \sqrt{N \sum_i y_i^2 - (\sum_i y_i)^2}} \quad (3)$$

where $N$ is the total number of observations, $x_i$ and $y_i$, $i=1, ..., N$, are observational variables. We used Pearson's correlation coefficient to calculate the correlation between predicted PoIRs and the corresponding value derived from ground truth. Furthermore, mean absolute percent error (m-APE), which is a assessing of prediction accuracy of a forecasting method, was herein used to measure the relative errors between the mean predicted PoIRs and the ground truth value on the test benchmark.

$$mAPE = \frac{1}{n} \sum_{i=1}^{n} |\frac{PoIR_{predicted} - PoIR_{ground-truth}}{PoIR_{ground-truth}}| \times 100\% \quad (3)$$

### *Use the value of HU*

The most straightforward way to segment desired lung regions was by aid of the value of HU as the threshold, which reflects the degree of X-ray absorption of different tissues. For instance, the HU value for lung parenchyma usually ranges from −800 to −500 and window of other soft tissue is from +100 to +300. This margin usually is solid enough to separate lung with other tissues. However, when there existed infected regions in lung, the HU value of lung parenchyma could extended to from -750 to 150 (based on our statistics on test benchmark in Fig 3). Therefore, the segmentation result with the threshold of HU usually is not typically accurate enough for clinical application. Alternatively, these labeled images could be used as the initial annotated samples in our study.

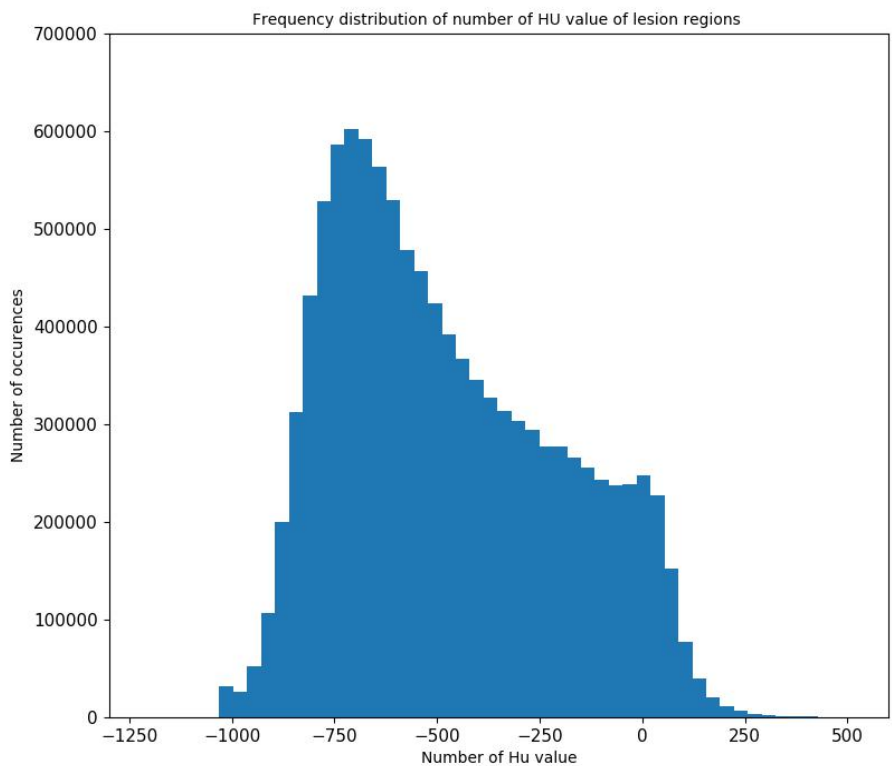

**Figure 3**. The distribution of HU value base on the ground truth of the proposed test benchmark. (Included some air tube and blood vessel tissue etc as they were hard to separation with lung structures). Most of the pixels were located within [–750, 150].

First, arithmetic progression of HU value (from –800 to 0 with increment of 50) was used as the threshold to segment the intact lung to achieve their corresponding masks. The segmented masks with maximum m-Dice (compared with the ground truth of intact lung) were used as the mask for intact lung. In the next step, use this mask for intact lung to minus the masks obtained with different HU value to obtain the mask of their difference values. The masks with difference in the maximum m-Dice (compared with the ground truth of infected regions) were used as the masks of infected regions.

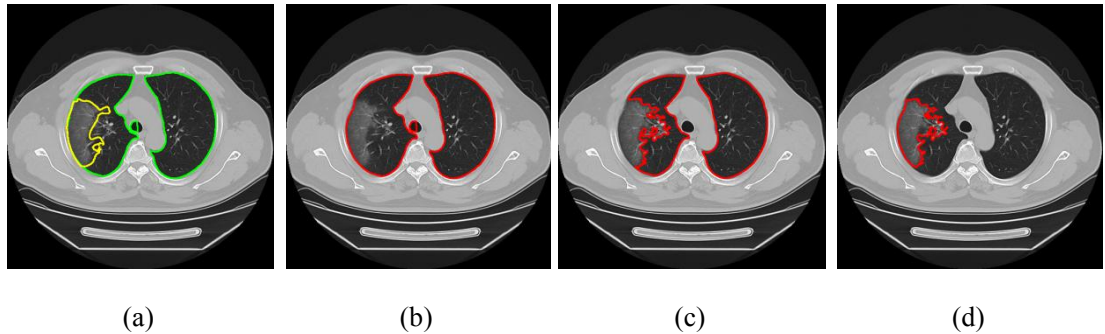

(a) (b) (c) (d)

**Figure 4.** (a) Ground truth of intact lung (green line) and infected regions (yellow line); (b) mask of lung obtained with HU = –200 (red line); (c) mask of lung obtained with HU = –750 (red line); (d) mask of infected regions was obtained by (b) minus (c) (red line).

### *Deep learning to segment the intact lung and the infectious regions*

As mentioned earlier, use of HU threshold cannot properly segmented the intact lung and infected regions. Those infected regions that had a close HU value with other soft tissues cannot be correctly differentiated. Therefore, deep learning techniques were utilized and evaluated in current study.

Training samples with detailed sketch of each infected region and intact lung are highly essential for developing the deep learning models. However, due to ambiguous edge between infected region and normal tissue, it was extremely timing-consuming to annotate thousands of lung CT images. The annotation result achieved by HU threshold was utilized for the preliminary samples. Then, two professional radiologists further manually contoured the intact lung and infected region based on the these "dirty" samples to generate final sample dataset for training and test.

*Network structure*

Two deep learning models were utilized: 2D UNet [15] (Fig. 5) and 2.5D UNet (Fig. 6). A two-dimensional (2D) deep learning models can well reflect the intra–slice information. However, they may neglect the inter–slice information and cannot fully leverage the spatial architecture of the three-dimensional (3D) slices of CT scans. On the other hand, 3D models [16, 17] suffer from tremendous increased parameters and the subsequent of hard to converge and overfitting especially for a limited number of training samples. Furthermore, due to the limitation of GPU memory, the original CT images had to be cropped or resized to small-sized cubes as the input for deep learning models. This crop or resize operation would either restrict the maximum inception regions or attenuate the resolution of original CT images. Therefore, a pseudo-3D segmentation or so called 2.5D UNet [18, 19] was used for evaluation purposes, in which the same UNet backbone (with expanded of network parameters) was used. In addition, three neighboring 2D slices were stacked as the inputs during training, so that the 2D network was able to detect a small range of 3D contexts each time. Three masks of image would be generated each time and the average value of segmentation maps would be used as the final masks for overlaps. The proposed two networks would be used both for the segmentation of intact lung and infected regions.

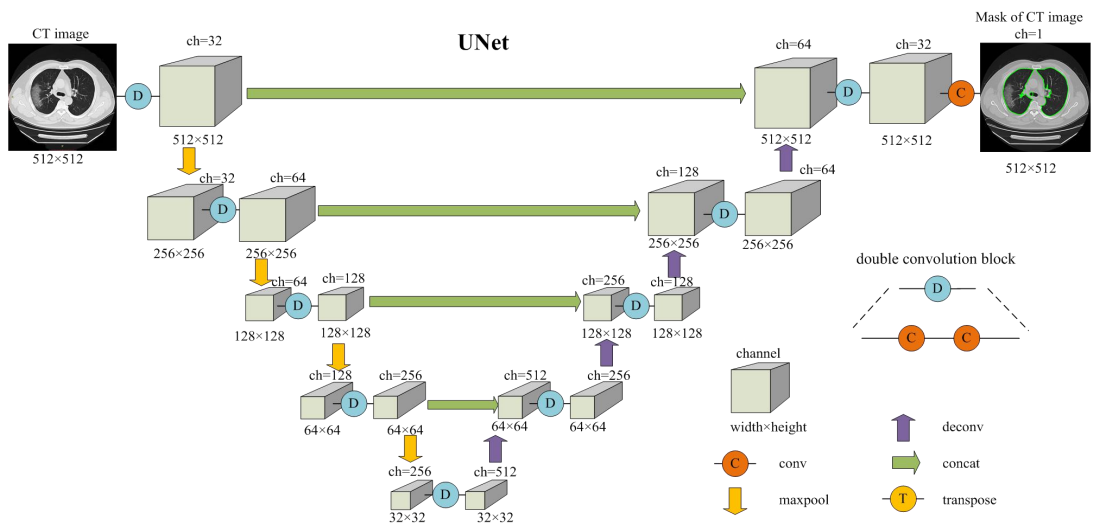

**Figure 5.** UNet model

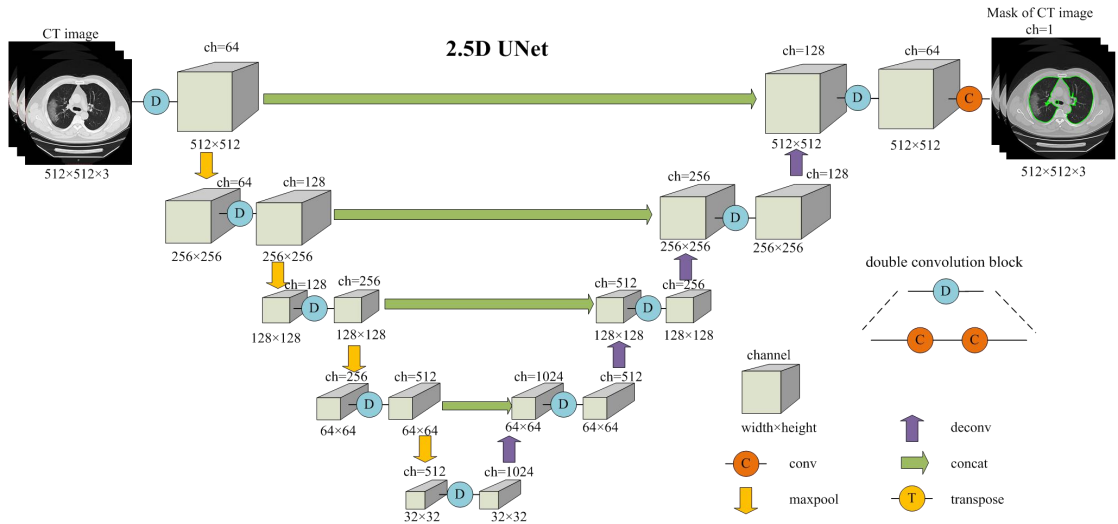

**Figure 6.** 2.5D UNet model.

## *Using a Classifier to further clarify infected regions*

It was observed that a certain number of blue-edged healthy structures, which had similar appearance as infected regions, were likely to be identified incorrectly. These kind of healthy tissues included air tube, blood vessel, and blur region of lung at the border etc as shown in Fig 7.

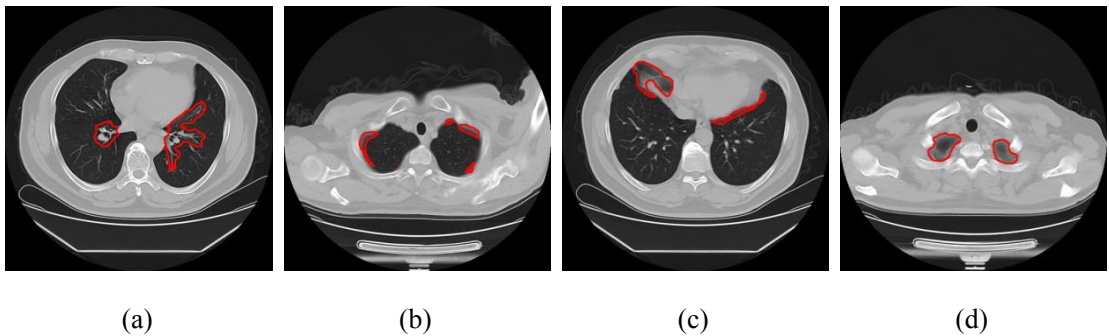

(a)          (b)          (c)          (d)

**Figure 7.** Regions in the red line are healthy regions (a) air tube and blood vessel; (b) (c) (d) healthy blur-edged regions at the border.

Therefore, a ResNet-18[20] based binary classifier (Fig. 8), was utilized after the segmentation models to further clarify whether an image patch belonged to infected regions or not. The masks corresponding to healthy regions were filtered. Compared with time-consuming pixel-level annotation on the blur-edged infected region, the training samples of this binary classification model could be relatively easily to be labeled and trained.

The candidate images from the output of segmentation model were firstly enclosed in a minimum circumscribed square bounding box. Then these image patches were used as the input data for the binary classifier to determine the valid existence of infected regions. Classical ResNet-18 network backbone was employed for image feature extraction part of the classifier. At the same time, generic data expansion mechanisms such as random clipping and left–right flipping were performed on specimens to increase the number of training samples and prevent data overfitting and improve the problem of generalization. The output of the convolution layer was flattened to a 256-dimensional feature vector, followed by three full-connection layers to export the final binary classification result.

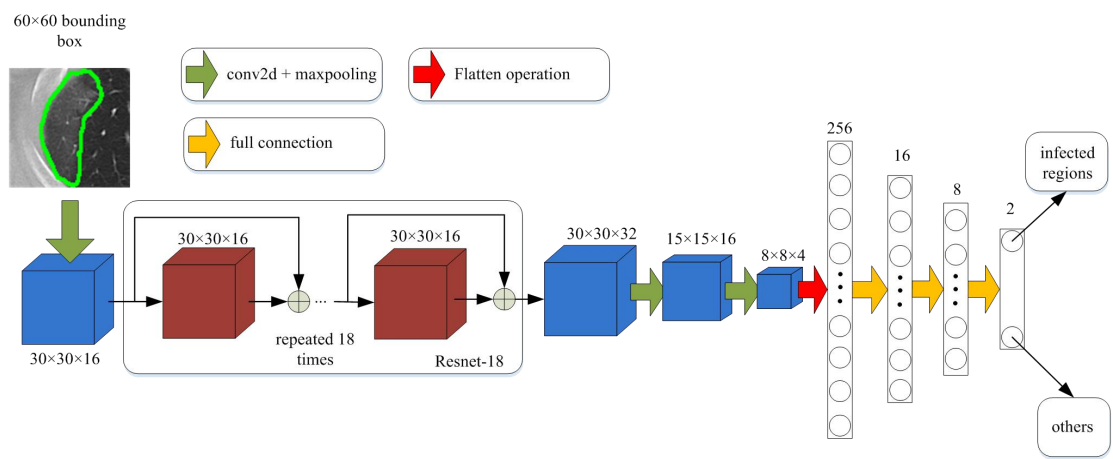

**Figure 8.** The network structure of ResNet18-based binary classification model

## Experiment results

Segment the intact lung and infected region with the threshold of HU as the initial annotated samples. Arithmetic progression value of HU (from −800 to 0 with the increment of 50) was used as the threshold to segment the intact lung directly to generate corresponding masks. Each outcome masks were evaluated on the test benchmark cases, as shown in Fig 9. The maximum m-Dice for intact lung and infected region was 0.921 (HU = −200) and 0.530 (HU = −750), respectively. The accuracy of m-Dice significantly decreased as many unrelated regions, e.g.

stomach, were wrongly segmented as lung when the threshold of HU was set to below –150.

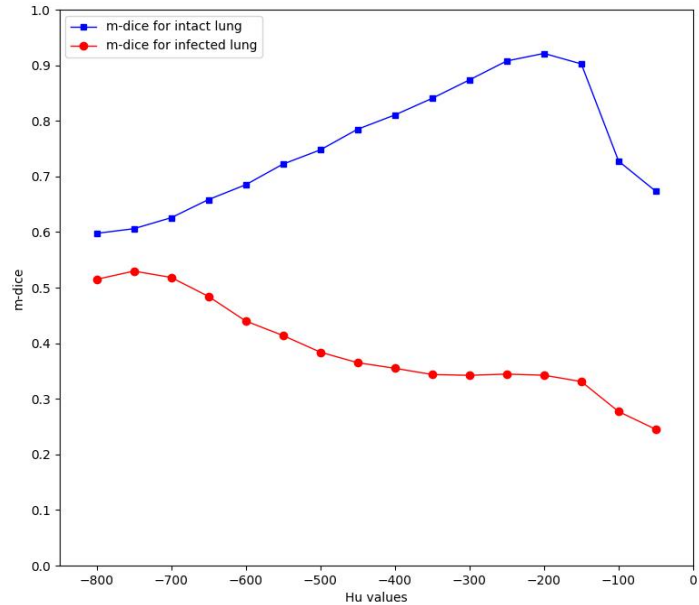

**Figure 9.** The maximum m-Dice for intact lung (HU = –200) and infected regions (HU = –750) could be achieved.

## *Generate mask for intact lung using deep learning models*

UNet and 2.5D UNet were utilized in the present study. For regions with light opacity, the threshold of HU along could achieve satisfactory segmentation results. However, for regions with high opacity, deep learning models could achieve obviously superior results. The m-Dice of UNet and 2.5D UNet of intact lung were 0.972 and 0.967, respectively. There was no significant difference between the two deep learning models.

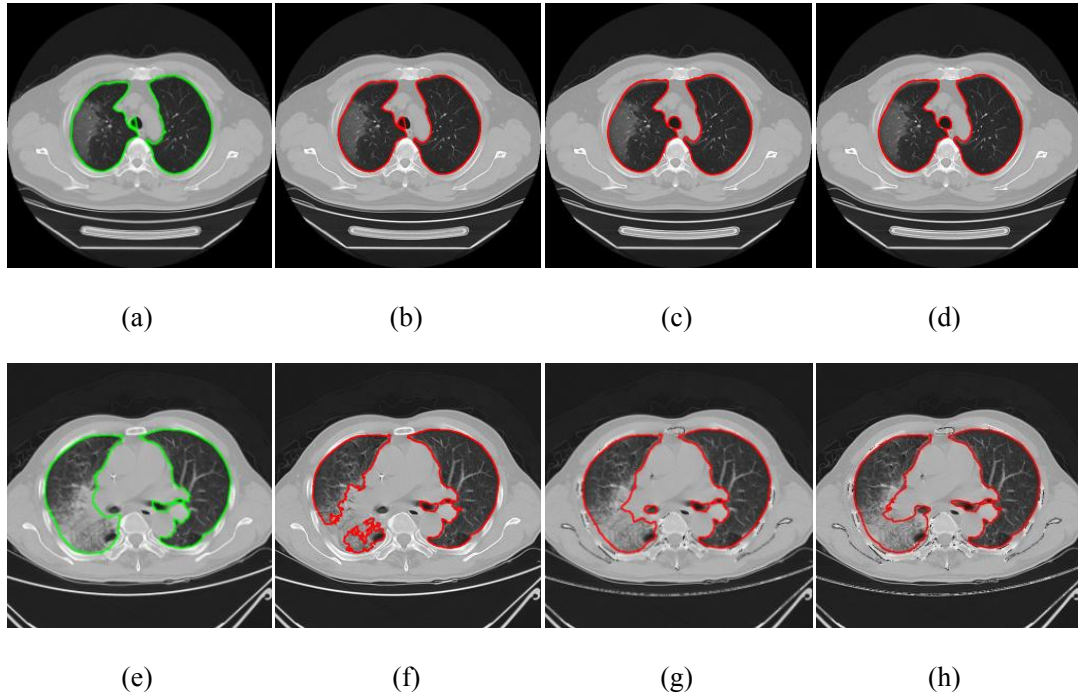

**Figure 10.** The first row of CT images were belongs to a case with light opacity, and the second row belongs to a case with high opacity. (a) Ground truth mask (green line); (b) Segmentation with HU (–200); (c) Segmentation with UNet; (d) Segmentation with 2.5D UNet; (e) Ground truth mask (green line); (f) Segmentation with HU (–200); (g) Segmentation with UNet; (h) Segmentation with 2.5D UNet.

## *Generate mask for infected regions*

The same UNet and 2.5D UNet deep learning models were evaluated. The m-Dice of UNet and 2.5D UNet for infected regions were 0.684 and 0.693, respectively. The latter model further concentrated on the inter-slice characteristics and demonstrated 1.3% improvement on the results. It was observed that most of the infected regions could be included in the output of the segmented masks. However, many blur-edged normal tissues were wrongly detected as infected regions, including air tube, blood vessel, stomach, and part of the border of lung etc. Therefore, a classifier was followed to further remove those unrelated regions.

*Performance of binary classifier*

The receiver operating characteristic curve (ROC) for the ResNet-18 based classifier was depicted in Figure 11. The value of area under curve (AUC) was 0.913, and when that valve equals to 0.45, the classification exhibited the best performance with accuracy of 93.8%. The m-Dice of UNet and 2.5D UNet of infected regions were 0.743 and 0.758 after the filtering this classification model, respectively. Compared with the results directly from the segmentation, the m-Dice improved 8.6% and 9.2%, respectively. The illustration of the segmentation result of infected regions and the summarized m-Dice for different methods were showed in Fig. 12 and Table 1.

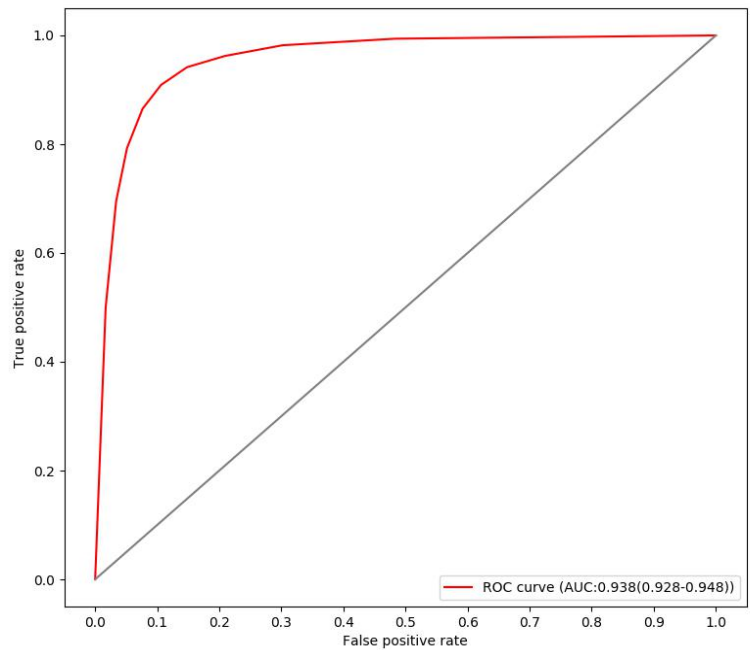

**Figure 11.** The ROC for the binary classifier.

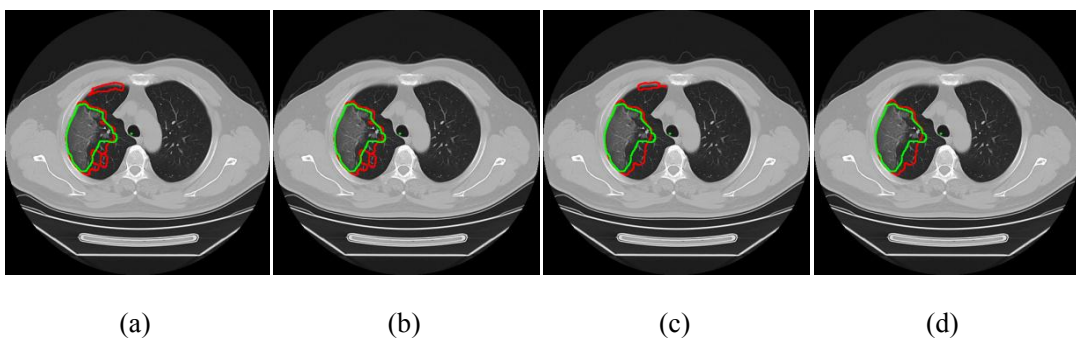

(a)          (b)          (c)          (d)

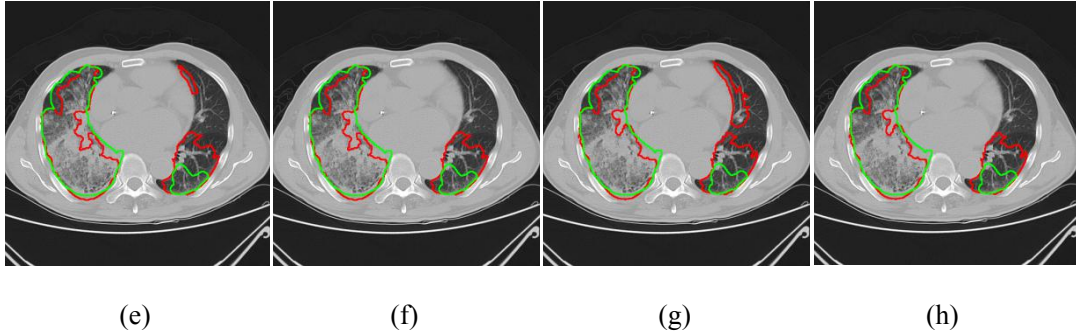

| | | (e) | (f) | (g) | (h) |

**Figure 12.** The first row of CT images belongs to a case with light opacity and the second row belongs to a case with high opacity. Ground truth mask marked with green line and predicted infected region marked in red line. (a) The prediction of UNet; (b) Further clarified by classifier; (c) The predictions of 2.5D UNet; (d) Further clarified by classifier; (e) The prediction of UNet; (f) Further clarified by classifier; (g) The prediction of 2.5D UNet; (h) Further clarified by classifier.

| | Methods | Mean Dice |
|---|---|---|
| Intact lung | HU only (= –200) | 0.941 |
| | UNet | **0.972** |
| | 2.5D UNet | 0.967 |
| Infected region | HU only (= –750) | 0.530 |
| | UNet | 0.684 |
| | 2.5D UNet | 0.693 |
| | UNet + classifier | 0.743 |
| | 2.5D UNet + classifier | **0.757** |

**Table 1.** Summarized m-Dice between predicted masks and the measurement derived from ground truth of our test benchmark.

With the aid of the masks of the intact lung and infected regions, the Pearson's correlation coefficient of the PoIRs could be achieved (0.961), which showed a very strong correlation between the predicted masks and those derived from ground truth. Furthermore, the m-APE of the PoIRs on test benchmark also could be obtained (11.7%), which indicated that the average relative errors between predicted PoIRs and the ground truth value was a lightly more than 10%.

**Discussion & conclusions**

With the rapid development of artificial intelligence technology, experiences of profession radiologists, such as the segmentation of medical images, could be solidified in the deep learning models to accomplish a quantitative analysis report. Several methods were developed to investigate the segmentation of intact lung and infected regions, including the threshold of HU, UNet, and 2.5D UNet. In addition, a fine-tuned classifier was followed to further remove those wrongly segmented healthy regions to improve the accuracy of outputs. We referred to the methodology of the design of nnUNet [21], which had achieved good results in many different medical segmentation tasks. They suggested that if the objected data is very anisotropic then a 2D UNet may actually be a better choice. For example, in the segmentation of pancreas, which was a blur-edged objective on the images as well, 2D network actually outperformed 3D counterparts.

As a matter of fact, the most challenge work in the calculation of the proportion of infected regions was the annotation of images, especially for the regions that affected by pneumonia. We utilize the intrinsic HU value of CT images to create the initial version of label images. Even though they were dirty samples, it would be much less effort for professional radiologist to further modify and improve on this first round version. Furthermore, the annotation of samples and the training of a fine-tuned binary classifier were much easier than the pixel-level of segmentation. Compared with direct result of the state-of-the-art segmentation algorithm, the classifier could improve the m-Dice of infected regions around 9%.

For the calculation of proportion of infected regions, the Pearson's correlation coefficient between predicted and the ground truth showed a strong correlation between them, which would be one of an objective indicator for monitoring the progress of one patient at a fixed interval. Furthermore, the m-APE showed promising outcomes for the reference for the decision of clinical physicians.

In the future, doctors can carry out quantitative analysis of the severity of COVID-19 patients with this model or combined with other clinical data such as blood oxygenation index. At the same time, they can compare the sequential CT scans of the same case to predict the prognosis and provide reliable basis for treatment.

However, this study had several limitations. In some cases, the segmentation models would possible identify healthy tissues together with valid infected regions and the following classifier could not remove this "valid" infected regions. Therefore, the corresponding mask in such scenario would be larger than the ground truth. Moreover, additional COVID-19 CT cases from different subtypes should be included to promote the accuracy of segmentation and classification. Some atypical infection signs, such as pleural effusions, cannot be distinguished with this model.

**Acknowledgements**

This study was supported by the Zhejiang province natural science fund for emergency research (LED20H190003)

This study was supported by the China national science and technology major project fund (20182X10101–001)

**Compliance with ethics guidelines**

All authors declare that they have no conflict of interest or financial conflicts to disclose.